%% file: main.tex
\documentclass[aps,prd,showpacs,superscriptaddress,nofootinbib,]{revtex4-2}

\usepackage{amsmath,amssymb}
\usepackage{graphicx}
\usepackage[linktoc=none]{hyperref}
\usepackage{xcolor}

\usepackage{orcidlink}

\definecolor{tab-blue}{RGB}{0, 107, 164}
\hypersetup{colorlinks=true,allcolors=tab-blue}

\begin{document}

\title{Testing Heavy Dark Matter Decay as the Origin of KM3-230213A}

\input{authors_new}

\begin{abstract}
This work explores the hypothesis that the ultra–high-energy neutrino event KM3-230213A originates from the decay of a heavy dark matter particle with a mass above PeV scale. 
The analysis exploits the deposited energy and arrival direction of the event as well as a complete detector Monte Carlo simulation to compute the expected signal distributions for different decay channels and assesses the relative contributions from Galactic and extragalactic dark matter. 
Assuming a dark matter origin of the event, we find that the preferred mass at 95\% C.L. is larger than about 100 PeV in all scenarios considered, with best-fit lifetimes in the range $10^{26}$--$10^{27}$ s. 
These preferred regions are in tension with existing bounds from other neutrino telescopes and gamma-ray observations. 
\end{abstract}

\maketitle

\section*{Introduction}

Dark matter (DM) constitutes a fundamental component of the current cosmological paradigm, accounting for the majority of the matter content of the Universe and playing a crucial role in structure formation and cosmic evolution. 
While there is compelling evidence for its existence from observations \cite{Wechsler:2018pic}, the nature of DM remains elusive, as no conclusive evidence for non-gravitational interactions has yet been observed despite extensive experimental efforts~\cite{Bertone:2004pz, Kahlhoefer:2017dnp, Planck:2018vyg, PerezdelosHeros:2020qyt, Billard:2021uyg}. 
Identifying the DM properties therefore presents a major open problem in contemporary physics~\cite{AlvesBatista:2021eeu}. 
In this context, indirect detection with high-energy neutrinos constitutes a powerful and complementary strategy to probe DM candidates through astrophysical observations, searching for potential signatures associated with DM annihilation or decay~\cite{Albert:2016emp, ANTARES:2016xuh, IceCube:2016oqp, IceCube:2016dgk, IceCube:2021xzo, ANTARES:2022aoa, IceCube:2022clp, Gozzini:2023nka, IceCube:2023ies, KM3NeT:2024xca}.

On February 13th, 2023, the KM3NeT/ARCA neutrino telescope observed an event compatible with a muon coming from $0.6^{\circ}$ above the horizon with an estimated energy of $120_{-60}^{+110}~{\rm PeV}$~\cite{KM3NeT:2025npi}.
Given its extreme energy and near-horizontal trajectory, the muon most likely originated from the interaction of a neutrino of even higher energy in the vicinity of the detector.
The origin of this neutrino, referred to as KM3-230213A, remains unknown, particularly as no comparable detections have been reported by either the IceCube Neutrino Observatory~\cite{IceCubeCollaborationSS:2025jbi}, Baikal-GVD~\cite{Baikal-GVD:2025kbe}, or Pierre Auger Observatory~\cite{PierreAuger:2023pjg}.
Proposed explanations include conventional astrophysical scenarios, such as Galactic and extragalactic accelerators~\cite{KM3NeT:2025ccp,KM3NeT:2025zmb,KM3NeT:2025lly,Fang:2025nzg,Filipovic:2025ulm,Crnogorcevic:2025vou,Yuan:2025zwe}, as well as cosmogenic neutrinos produced through interactions of ultra–high-energy (UHE) cosmic rays with the cosmic microwave background~\cite{KM3NeT:2025vut,Zhang:2025abk,Boxi:2025ony,Das:2025vqd}.
In addition, the event has stimulated interest in more exotic interpretations invoking physics beyond the Standard Model~\cite{KM3NeT:2025mfl, Dvali:2025ktz, Klipfel:2025jql, Jiang:2025blz, Alves:2025xul, Narita:2025udw, Brdar:2025azm, Boccia:2025hpm, Amelino-Camelia:2025lqn, Bertolez-Martinez:2025trs, Airoldi:2025opo, Anchordoqui:2025xug, Dev:2025czz, Farzan:2025ydi, Baker:2025cff, Choi:2025hqt, Sakharov:2025oev}, including the possibility that KM3-230213A originates from the decay of heavy DM particles~\cite{Barman:2025hoz, Jho:2025gaf, Aloisio:2025nts, Kohri:2025bsn, Borah:2025igh, Khan:2025gxs, Murase:2025uwv}. 
Indeed, heavy DM candidates decaying on cosmological scales emerge in several theoretical frameworks~\cite{Chang:1996vw, Greene:1997ge, Chung:1998zb, Benakli:1998ut, Kolb:1998ki, Coriano:2001mg, Garny:2015sjg, Kannike:2016jfs, Kolb:2017jvz, Kim:2019udq, Dudas:2020sbq, Hambye:2020lvy, Ling:2021zlj, Allahverdi:2023nov}.

In this work, we assess the compatibility of the KM3-230213A event with the hypothesis of decaying DM with mass larger than $10~{\rm PeV}$.
Previous studies~\cite{Barman:2025hoz, Jho:2025gaf, Aloisio:2025nts, Kohri:2025bsn, Borah:2025igh, Khan:2025gxs, Murase:2025uwv} inferred the DM mass $m_{\rm DM}$ and lifetime $\tau_{\rm DM}$ by comparing the reported neutrino flux in Ref.~\cite{KM3NeT:2025npi}, derived assuming an $E^{-2}$ spectrum, with DM-induced neutrino spectra that exhibit substantially different shapes.
In contrast, our analysis relies on the full simulation chain of KM3NeT, which allows us to propagate the neutrino spectra from DM decay through the detector response and consistently estimate the corresponding deposited energy and angular distributions. 
We use this framework to infer $m_{\rm DM}$ and $\tau_{\rm DM}$ for several two-body decay channels and to quantify the relative DM contributions from Galactic and extragalactic components.
Finally, we compare our results with existing constraints from other neutrino telescopes and gamma-ray observatories.

\section*{Detection technique}

The passage of relativistic charged particles through a transparent dielectric causes the emission of Cherenkov light, and lays the groundwork for neutrino detection in very-large volume telescopes.
These instruments, built as photomultiplier arrays, detect neutrinos from the Cherenkov light of their daughter charged particles produced in interactions with matter in or in the vicinity of the detector.
This analysis is based on a sample of track-like events originating primarily from charged-current muon-neutrino interactions, in which the outgoing muons can propagate over long distances through the detector medium.
Cherenkov photons are emitted along a cone around the muon trajectory, making it possible to estimate the muon direction from the arrival times and positions of the detected photons.
The number of detected Cherenkov photons is correlated with the muon energy, which in turn is correlated with the parent neutrino energy, allowing an estimation of the neutrino energy from the observed Cherenkov light.

The KM3NeT neutrino telescopes ARCA (Astroparticle Research with Cosmics in the Abyss) and ORCA (Oscillation Research with Cosmics in the Abyss) use the Mediterranean Sea water as a detection medium~\cite{LOI}.
The Cherenkov light produced in or in the vicinity of the detector is observed by Digital Optical Modules (DOMs) joined in vertical detection lines.
Each DOM hosts 31 photomultiplier tubes (PMTs), along with sensors to determine their position and orientation~\cite{KM3NeT_DOM}.
The most abundant background, caused by muons from collisions in the atmosphere of cosmic rays, not connected with a cosmic neutrino interaction, is suppressed by selecting only events that reach the detector after crossing the Earth.
In its final design, KM3NeT/ARCA will be a cubic-kilometer detector, sensitive to low fluxes of cosmic neutrinos. 
It has a vertical spacing of 36 m between the DOMs and a horizontal spacing of about 90 m between the detection units, and lies at a latitude of $36^\circ 16'$ N at a depth of 3.5 km.
During construction, partial detector configurations started recording data promptly after connection of each unit. 
The multi-PMT layout of the KM3NeT DOMs contributes to the excellent pointing precision of the instrument~\cite{KM3NeT:2024paj}.
Additionally, water has optimal optical properties: due to its large scattering length, most of the emitted Cherenkov photons reach the PMTs without having scattered.

The reconstruction of the track direction uses a maximum likelihood algorithm that includes the measured times and positions of the first hits recorded with the PMTs.
The deposited energy is reconstructed using a one-dimensional likelihood that includes the spatial distribution of hit and non-hit PMTs around the reconstructed trajectory.
This observable differs from that used in Ref.~\cite{KM3NeT:2025npi}, where the energy of the muon prior to entering the detector was reported.

This analysis follows the event selection strategy described in Ref.~\cite{KM3NeT:2025npi} to isolate very bright track-like signatures crossing the detector.
High quality reconstructed events are selected by requiring the track length to be longer than 250 meters and the log-likelihood ratio observable quantifying the reconstruction quality to be larger than 500.
We also apply a cut to eliminate the downgoing atmospheric muon background by requiring that the estimated overburden, based on the track direction, is longer than 30 km. 
At this overburden, the atmospheric muon contribution is subdominant to all other backgrounds considered in this analysis and can be safely neglected.

In this work we use the data collected with the ARCA21 configuration from September 25th, 2022 to September 11th, 2023, corresponding to a lifetime of 286 days.

\section*{Analysis} 

We conduct a frequentist analysis using a binned likelihood function with two free parameters: the mass $m_{\rm DM}$ and the lifetime $\tau_{\rm DM}$ of DM particles.
The simulated detector response is convolved with different neutrino flux models to compute the expected number of events. 

The neutrino flux associated with DM decays receives contributions from both the Galactic and extragalactic distributions. 
The Galactic differential flux of neutrinos and antineutrinos of flavor $\alpha$ per neutrino energy $E_{\nu}$ and solid angle $\Omega$ can be written as
\begin{equation}\label{eq:galDM}
    \frac{\mathrm{d}\Phi^\text{Gal}_{\nu_{\alpha}+\bar{\nu}_{\alpha}}}{\mathrm{d}E_\nu \mathrm{d}\Omega}
    =\frac{1}{4\pi \, m_\text{DM}\, \tau_\text{DM}}
    \frac{\mathrm{d}N_\alpha}{\mathrm{d}E_\nu}
    \int_0^\infty \mathrm{d}s\,\rho_\text{DM}(r)\,,
\end{equation}
where $\rho_{\rm DM}$ is the DM density profile with radial distance $r = (s^2 + R_\odot^2 - 2sR_\odot \cos l \cos b)^{1/2}$ in the galactocentric coordinate system.
The $l$ and $b$ are Galactic longitude and latitude, respectively, and $R_\odot = 8.178$~kpc is the Sun's galactocentric distance~\cite{Gravity:2019nxk}.
The integral is performed over the line-of-sight distance $s$ measured from the Earth.
This flux depends on the energy distribution $\mathrm{d}N_\alpha/\mathrm{d}E_\nu$ of neutrinos of flavor $\alpha$ produced in the decay of an individual DM particle with a specific decay channel. 
In this analysis, we employ the neutrino spectra computed with the public code \texttt{HDMSpectra}~\cite{Bauer:2020jay}, which accounts for the electroweak radiative corrections, that are important for $m_{\rm DM} \gg 100~{\rm GeV}$, as well as for the subsequent particle cascades triggered by the primary decay.
We test three benchmark scenarios where DM particles decay to one channel with a 100\% branching ratio at a time: neutrinophilic DM particles (${\rm DM} \to \nu \overline{\nu}$) with diagonal flavor couplings, leptophilic DM particles decaying into tau leptons (${\rm DM} \to \tau \overline{\tau}$), and hadrophilic DM particles decaying into bottom quarks (${\rm DM} \to b \overline{b}$).

The angular distribution of the neutrino arrival directions is encoded in the DM density profile of the Milky Way halo $\rho_\text{DM}(r)$.
Here, we consider both the Navarro--Frenk--White (NFW) ~\cite{Navarro:1995iw} and the Burkert~\cite{Burkert:1995yz} profiles, which are representatives of cuspy and cored halo models, respectively~\cite{deBlok:2009sp}.
Specifically, the NFW parametrization takes the expression
\begin{equation}
    \rho^{\rm NFW}_\mathrm{DM}(r)=\frac{\rho_s}{(r/r_s)\left(1+r/r_s\right)^2}\,,
\end{equation}
where we take the normalization and the scale radius to be $\rho_s = 0.23~{\rm GeV \, cm^{-3}}$ and $r_s = 25~{\rm kpc}$, respectively.
On the other hand, the Burkert profile is defined as
\begin{equation}
    \rho^{\text{Bur}}_{\rm DM}(r)=\frac{\rho_{c}R_{c}^3}{(r+R_{c})(r^2+R_{c}^2)}\,,
\end{equation}
with the core radius and the central density being $R_c = 25~{\rm kpc}$ and $\rho_c = 0.59~{\rm GeV \, cm^{-3}}$, respectively. 
The profile parameters provide the reference value for the local DM density of $0.4~{\rm GeV \,cm^{-3}}$ in agreement with the latest estimates~\cite{Benito:2019ngh, Benito:2020lgu}.
With these parameters, the NFW profile yields a larger contribution for angular distances from the Galactic center below $30^\circ$, while the Burkert profile dominates at larger angles.
We note that DM substructures (clumps) are not included in our analysis since their impact is negligible in decaying DM scenarios.

The extragalactic component can be expressed as
\begin{equation}\label{eq:extragalDM}
    \frac{\mathrm{d}\Phi^\text{egal}_{\nu_{\alpha}+\bar{\nu}_{\alpha}}}{\mathrm{d}E_\nu \mathrm{d}\Omega}=\frac{\Omega_\text{DM}\, \rho_{crit}}{4\pi\, m_\text{DM}\, \tau_\text{DM}} \int_0^\infty \frac{\mathrm{d} z}{H(z)} \frac{\mathrm{d}N_\alpha}{\mathrm{d}E_\nu}\bigg{\vert}_{E^\prime_\nu=E_\nu(1+z)} \,,
\end{equation}
where $\rho_{crit} \simeq 5.5 \times 10^{-6}~{\rm GeV \, cm^{-3}}$ is the critical density of the Universe, $\Omega_\text{DM} = 0.264$ encodes the density of the DM particles in the Universe, $z$ is the cosmological redshift and $H(z)$ is the Hubble expansion rate as inferred by the latest Planck measurements of the cosmological parameters~\cite{Planck:2018vyg}.
In contrast to the Galactic component, the extragalactic flux in Eq.~\eqref{eq:extragalDM} is isotropic and has a modified energy distribution given by the integral over the redshift.
Specifically, the energy distribution $\mathrm{d}N_\alpha/\mathrm{d}E_\nu$ is computed at the energy $E^\prime_\nu$ which corresponds to the energy $E_\nu$ after propagation. 
Propagation of neutrinos over astrophysical distances leads to the averaging out of flavor oscillation probabilities.
These probabilities were calculated according to the latest global analysis of the neutrino oscillation parameters~\cite{Capozzi:2025wyn} which were also used to calculate the flux of neutrinos of a given flavor $\alpha$ reaching the Earth.
Contrary to gamma rays, the absorption of neutrinos during their propagation is negligible and was not considered for calculating their fluxes.

For the background, we account for a contribution associated with the production of neutrinos in the atmosphere, of which prompt production dominates at the energies relevant for this analysis.
To include this contribution we use the ERS08 model~\cite{Enberg:2008te}, from which we expect $3\times10^{-4}$ events passing the event selection. In addition, we assume an astrophysical diffuse neutrino flux component that we model using a single power law energy spectrum.
To do so, we extrapolate the flux measured by IceCube from two independent samples, namely high-energy starting events (HESE)~\cite{IceCube:2020wum} and through-going events (NST)~\cite{Abbasi:2021qfz}. The estimated fluxes for these two samples widely differ for energies above 1 PeV.
The expected number of events that pass the selection criteria ranges from $4\times10^{-3}$ to $24\times10^{-3}$, being at least an order of magnitude larger than the atmospheric contributions.
In this analysis, we also test a hypothesis in which no astrophysical background is introduced, which would mimic a scenario in which there is a cut-off in the PeV energy regime.

\begin{figure}[t!]
\centering
\includegraphics[width=0.45\linewidth]{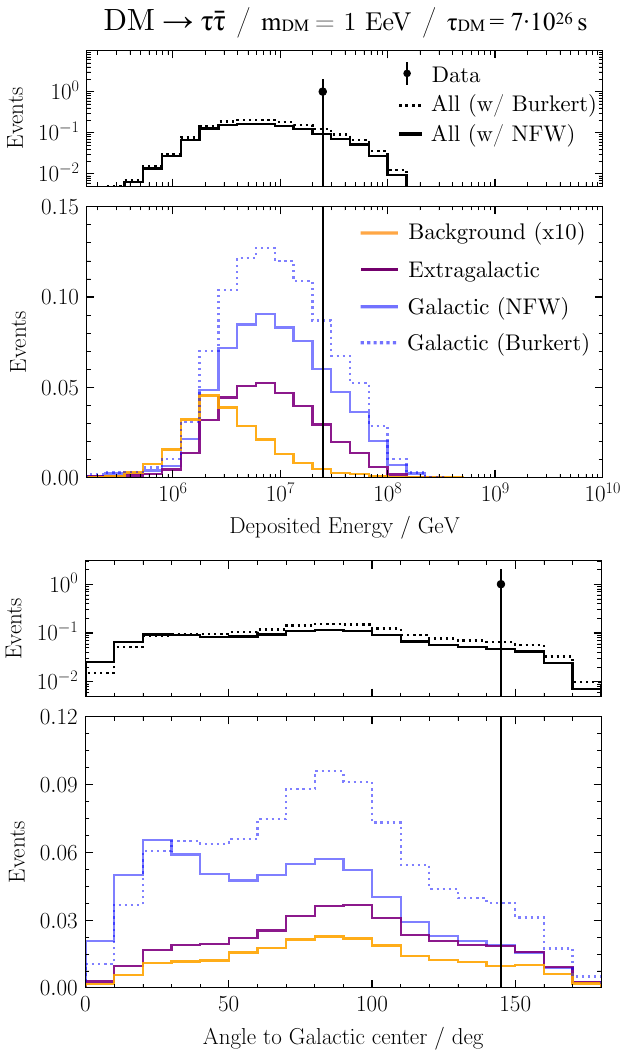}
\caption{
Expected distributions of the deposited energy (top) and of the angle with respect to the Galactic center (bottom) assuming DM decaying to $\tau\bar{\tau}$ with $m_{\rm DM}=1$~EeV and $\tau_{\rm DM}=7\times10^{26}$~s.
The upper panels show in logarithmic scale the expected number of events summing all the flux components, assuming a NFW (solid) or Burkert (dashed) profile for the Galactic contribution.
The lower panels show in linear scale the contribution from each flux component: Galactic, assuming NFW (solid blue) and Burkert (dashed blue) profiles, extragalactic (purple), and background scaled up by a factor of 10 (orange). 
Data points are black markers with error bars representing the Poissonian statistical error (only one survives the selection, KM3-230213A).
}
\label{fig:reco_dist}
\end{figure}

In Figure~\ref{fig:reco_dist} we show the expected distribution of the deposited energy and of the angle to the Galactic center for simulated events passing the selection criteria.
For the signal we assumed the best-fit parameters for DM decay to $\tau\bar{\tau}$ (see Table~\ref{tab:bestfit}) and for the background we include both atmospheric neutrinos and cosmic neutrinos using the extrapolated flux from NST.  
In the data, only one event survives the selection, KM3-230213A. 
The contributions from the Galactic and extragalactic components in the region where the event lies are at the same level for a NFW profile, as indicated by the expected angular distribution.
Assuming a Burkert profile, the Galactic contribution is $\sim70\%$ at that direction, as the corresponding angular distribution is less peaked toward the Galactic center.

%==============================
\section*{Results} 

A best-fit point in the $\tau_{\rm DM}$--$m_{\rm DM}$ parameter space is found for each decay channel and astrophysical background hypothesis from our likelihood analysis.
The confidence intervals are constructed using Feldman-Cousins procedure~\cite{Feldman:1997qc}.
We present these results assuming a NFW profile for the Galactic component but adopting a Burkert profile yields very similar confidence intervals.
The best-fit points, $m_{\rm DM}^{\textrm{best-fit}}$ and $\tau_{\rm DM}^{\textrm{best-fit}}$, for each decay channel are shown in Table~\ref{tab:bestfit}. 

\begin{table}[t!]
\centering
\begin{tabular}{c|c|c}
Decay channel & $m_{\rm DM}^{\textrm{best-fit}}$/EeV & $\tau_{\rm DM}^{\textrm{best-fit}}/10^{26}$s \\
\hline
$\nu\bar{\nu}$ & 0.2 & 4 \\
$\tau\bar{\tau}$ & 1 & 7 \\
$b\bar{b}$ & 100 & 4 \\
\end{tabular}
\caption{Best-fit values of $m_{\rm DM}$ and $\tau_{\rm DM}$ for the different decay channels considered in this analysis.}
\label{tab:bestfit}
\end{table}

\begin{figure}[t!]
\centering
\includegraphics[width=0.5\linewidth]{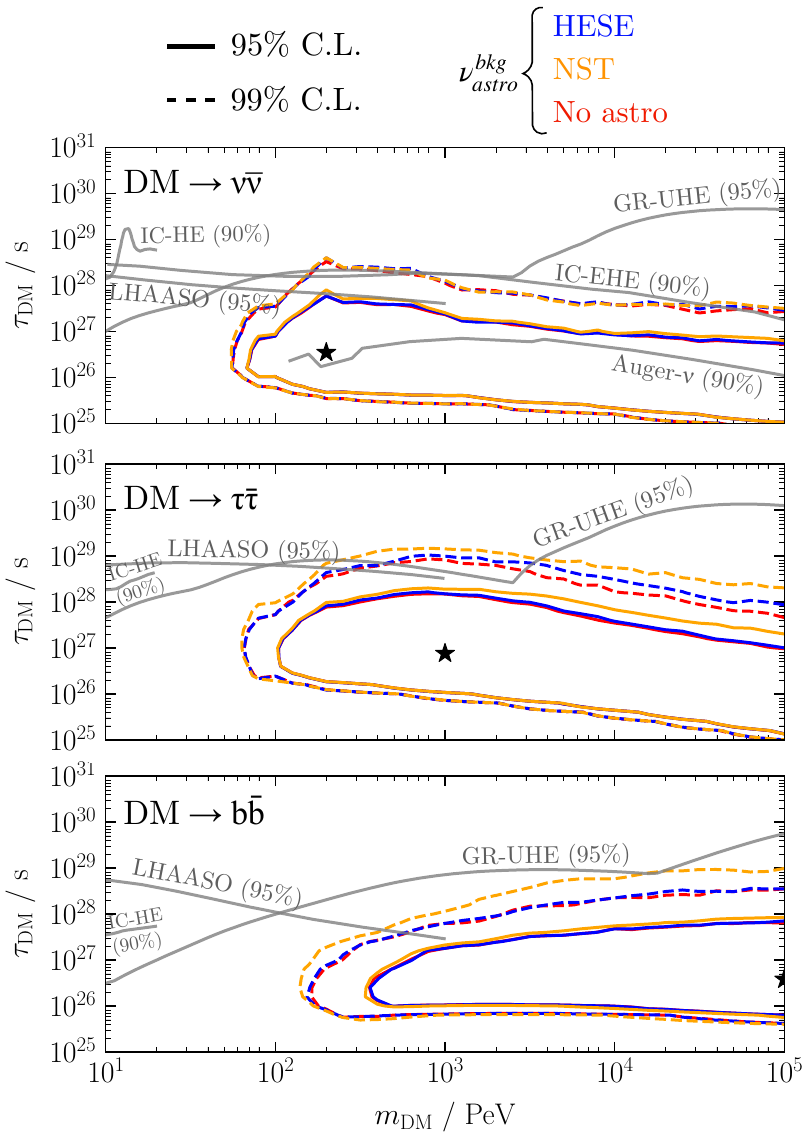}
\caption{
The Feldman–Cousins 95\% and 99\% C.L. contours (solid and dashed lines, respectively) for DM decaying into $\nu\bar{\nu}$ (top), $\tau\bar{\tau}$ (middle) and $b\bar{b}$ (bottom) assuming a NFW Galactic profile and different astrophysical diffuse neutrino flux scenarios: HESE (blue), NST (orange), no astrophysical component (red). 
The star markers represent the best-fit point. 
The gray lines indicate world-data lower limits (parentheses indicate the reported C.L.): IC-HE (IceCube's high energy starting events~\cite{IceCube:2022clp}), GR-UHE (UHE gamma-ray data~\cite{Chianese:2026cfz}), LHAASO~\cite{LHAASO:2022yxw}, IC-EHE (IceCube's extremely high energy analysis~\cite{Arguelles:2022nbl}), and Auger-$\nu$ (Auger's neutrino analysis~\cite{Arguelles:2022nbl}).
}
\label{fig:limits}
\end{figure}

\begin{figure}[t!]
\centering
\includegraphics[width=0.5\linewidth]{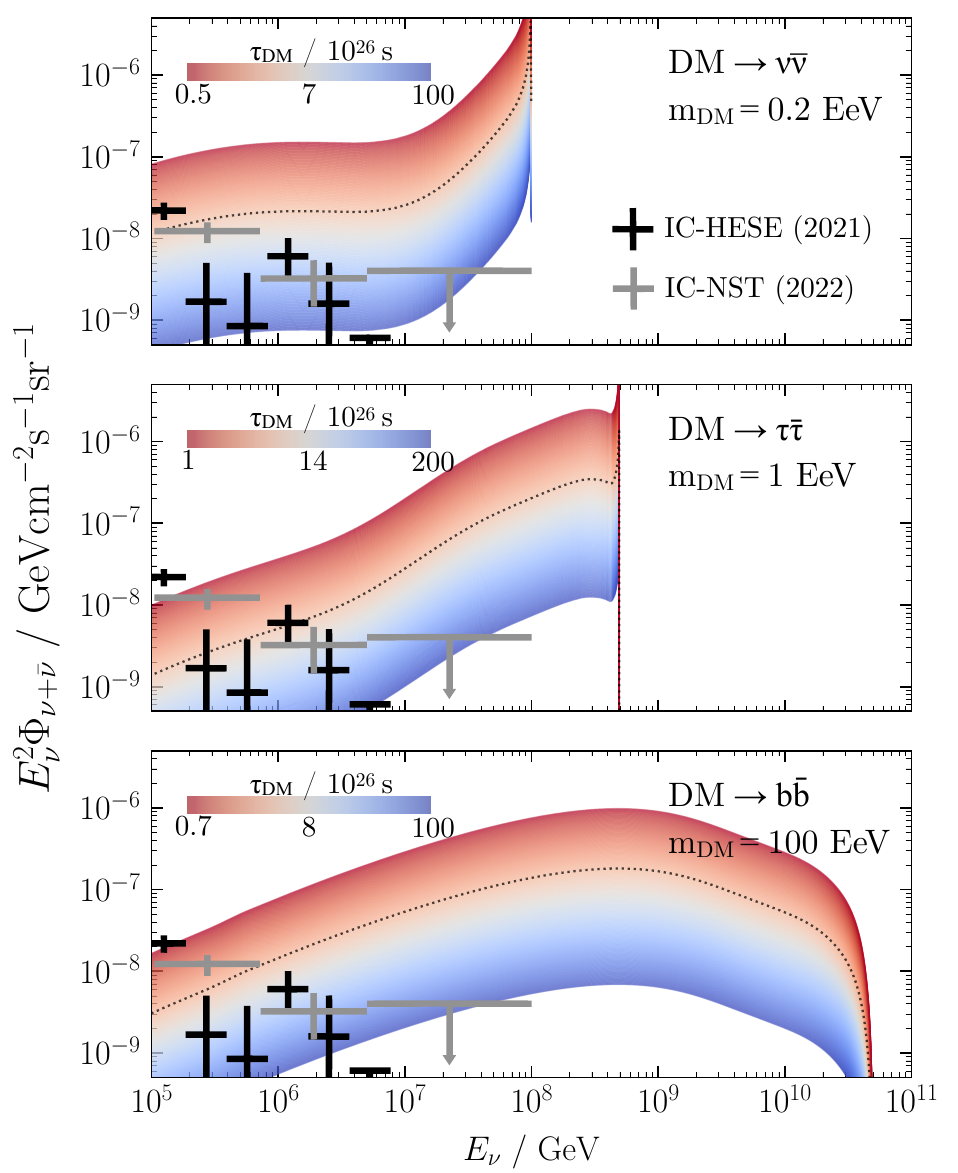}
\caption{
Each panel shows the neutrino energy flux per flavor as a function of neutrino energy for the different DM decay channels: $\nu{\bar\nu}$ (top), $\tau{\bar\tau}$ (middle) and $b{\bar b}$ (bottom).
The value of the DM mass matches the best-fit hypothesis in each channel, defined as $m_{\textrm{DM}}^{\textrm{best-fit}}$ (see Table~\ref{tab:bestfit}). 
The color map indicates the value of $\tau_{\textrm{DM}}$, which ranges in log-scale from the lower to the upper bounds at 95\% C.L. for a fixed $m_{\textrm{DM}}^{\textrm{best-fit}}$.
The dashed lines represent the flux for the best-fit $\tau_{\textrm{DM}}$.
The data points represent the IceCube's measurements of the astrophysical diffuse flux with NST~\cite{Abbasi:2021qfz} and HESE~\cite{IceCube:2020wum} samples.
}
\label{fig:flux}
\end{figure}

In Figure~\ref{fig:limits} we show the Feldman–Cousins 95\% and 99\% C.L. contours for each decay channel and background hypothesis assuming an NFW profile.
The preferred regions behave similarly for the three astrophysical background hypotheses.
As expected, larger contours are obtained using the NST spectrum, which predicts the largest background contribution. The 95\% contours have lower mass bounds of 300, 100, and 80 PeV for the $b\bar{b}$, $\tau\bar{\tau}$, and $\nu\bar{\nu}$ channels, respectively, and extend up to masses of order 100 EeV.
This extension to very high masses arises from the energy tails for $E_{\nu} \ll m_{\rm DM}$ of the decay spectra in all three channels (see Figure~\ref{fig:flux}), which still allow the KM3-230213A event to be accommodated even for very large $m_{\rm DM}$.
Constraining this high-mass region is further limited by the fact that only a single event survives the selection and by the lack of detector simulations beyond 100 EeV neutrino energy, which sets the upper bound of the scanned range.

The results obtained in this work are compared with bounds from gamma-ray experiments~\cite{Esmaili:2015xpa, Kalashev:2016cre, Ishiwata:2019aet, Chianese:2021jke, LHAASO:2022yxw, PierreAuger:2022jyk, PierreAuger:2022ubv, Das:2023wtk, Leung:2023gwp, Munbodh:2024ast, Chianese:2026cfz}, which can also be obtained for the neutrinophilic channel due to the gamma-ray production via the electroweak corrections~\cite{Berezinsky:2002hq}.
For $m_{\rm DM} \lesssim 100~{\rm PeV}$, the strongest gamma-ray constraints on the lifetime of heavy DM particles are placed by LHAASO-KM2A~\cite{LHAASO:2022yxw}; for higher DM masses, stringent constraints are instead placed by the non-observation of UHE gamma rays by different air-shower experiments~\cite{CASA-MIA:1997tns, KASCADE:2005ynk, KASCADEGrande:2017vwf, TelescopeArray:2018rbt, PierreAuger:2022aty, PierreAuger:2022uwd, PierreAuger:2024ayl, PierreAuger:2025jwt}.
The UHE gamma-ray limits in Ref.~\cite{Chianese:2026cfz} are obtained by comparing the primary gamma-ray emissions resulting from the prompt DM decays in the Galactic halo (assuming the same density profile adopted in this analysis) with the most recent upper limits on the isotropic UHE gamma-ray flux, accounting for the field of view and the geometric acceptance of each gamma-ray observatory.
The preferred regions in the three decay channels as shown in Figure~\ref{fig:limits} are excluded by the global gamma-ray lower bound at 95\% C.L. on the DM lifetime derived in Ref.~\cite{Chianese:2026cfz}. 

Other analyses have also looked for heavy DM signatures and can be compared with this analysis.
For the $\nu\bar{\nu}$ decay channel, the non-observation of UHE neutrinos by IceCube and Pierre Auger Observatory was recasted into lower limits on $\tau_{\textrm{DM}}$ for different masses~\cite{Arguelles:2022nbl}.
These experiments show smaller tension with the results presented here compared to gamma-ray limits but certain preferred regions are also excluded (see Figure~\ref{fig:limits}).
DM decay channels other than $\nu{\bar\nu}$ have not been studied with these samples.

There have also been dedicated searches of heavy DM particles by IceCube using the HESE sample, IC-HE~\cite{IceCube:2022clp}.
However, the reported limits only reach masses of 20 PeV in those searches.
Nevertheless, in Figure~\ref{fig:flux} it is shown that all the best-fit scenarios, for which the $\tau_{\textrm{DM}}$ is $\mathcal{O}(10^{26}-10^{27}~{\rm s})$, predict DM-induced neutrino fluxes that are in tension with the diffuse spectrum measured by IceCube with the NST and HESE samples.
On the other hand, for $\tau_{\textrm{DM}}\sim10^{28}~{\rm s}$ the predicted fluxes are significantly smaller and the tension with IceCube's diffuse measurements would be reduced.

\section*{Conclusions} 

In this paper the possibility is investigated that the KM3-230213A event originates from neutrinos produced in the decay of heavy DM particles.
Data collected with the ARCA21 detector configuration were analyzed by means of a binned likelihood approach, based on full Monte Carlo simulations of the detector response. 
Preferred regions for the DM mass and lifetime assuming different decay channels were identified and have been compared with current constraints from high-energy gamma ray and neutrino experiments.

A comparison of our result with existing experimental limits and with the expected Galactic and extragalactic contributions to the DM density profile yields two main insights.
First, only limited portions of the preferred parameter space remain compatible with existing gamma-ray constraints and do not overshoot the measured diffuse cosmic neutrino flux reported by other neutrino telescopes. 
Secondly, assuming a DM hypothesis, no definitive conclusion can be drawn regarding a Galactic or extragalactic origin of the KM3-230213A event, as the reconstructed angular distribution indicates that both contributions are comparable in the direction of the observed event  either for NFW or Burkert profiles.
Finally, this work constitutes the first analysis consistently convolving the neutrino flux from heavy DM decay with the KM3NeT/ARCA detector response.
Extending this framework to include data from additional neutrino telescopes will be needed to further constrain heavy DM properties.

\section*{Acknowledgments}

The authors acknowledge the financial support of:
%INFRADEV
KM3NeT-INFRADEV2 project, funded by the European Union Horizon Europe Research and Innovation Programme under grant agreement No 101079679;
%Belgium
Funds for Scientific Research (FRS-FNRS), Francqui foundation, BAEF foundation;
%Czeck
Czech Science Foundation (GAČR 24-12702S);
%France
Agence Nationale de la Recherche (contract ANR-15-CE31-0020), Centre National de la Recherche Scientifique (CNRS), Commission Europ\'eenne (FEDER fund and Marie Curie Program), LabEx UnivEarthS (ANR-10-LABX-0023 and ANR-18-IDEX-0001), Paris \^Ile-de-France Region, Normandy Region (Alpha, Blue-waves and Neptune),
%For the CPER
The Provence-Alpes-Côte d'Azur Delegation for Research and Innovation (DRARI), the Provence-Alpes-Côte d'Azur region, the Bouches-du-Rhône Departmental Council, the Metropolis of Aix-Marseille Provence and the City of Marseille through the CPER 2021-2027 NEUMED project,
%For IN2P3
The CNRS Institut National de Physique Nucléaire et de Physique des Particules (IN2P3);
%Georgia
Shota Rustaveli National Science Foundation of Georgia (SRNSFG, FR-22-13708), Georgia;
%Germany (Max Planck Inst.)
This research was funded by the European Union (ERC MuSES project No 101142396); This work was supported by the European Research Council, ERC starting grant MessMapp, under contract No. 949555.
%Greece
The General Secretariat of Research and Innovation (GSRI), Greece;
%Italy
Istituto Nazionale di Fisica Nucleare (INFN) and Ministero dell’Universit{\`a} e della Ricerca (MUR), KM3NeT4RR MUR Project National Recovery and Resilience Plan (NRRP), Mission 4 Component 2 Investment 3.1, Funded by the European Union – NextGenerationEU,CUP I57G21000040001, Concession Decree MUR No. n. Prot. 123 del 21/06/2022;
M. Chianese acknowledges the support by the research project TAsP (Theoretical Astroparticle Physics) funded by the INFN;
%Morocco
Ministry of Higher Education, Scientific Research and Innovation, Morocco, and the Arab Fund for Economic and Social Development, Kuwait;
%The Netherlands
Nederlandse organisatie voor Wetenschappelijk Onderzoek (NWO), the Netherlands;
%Poland
The grant “AstroCeNT: Particle Astrophysics Science and Technology Centre”, carried out within the International Research Agendas programme of the Foundation for Polish Science financed by the European Union under the European Regional Development Fund; The program: “Excellence initiative-research university” for the AGH University in Krakow; The ARTIQ project: UMO-2021/01/2/ST6/00004 and ARTIQ/0004/2021;
%Romania
Ministry of Education and Scientific Research, Romania;
%Slovak Republic
Slovak Research and Development Agency under Contract No. APVV-22-0413; Ministry of Education, Research, Development and Youth of the Slovak Republic;
%Spain
MICIU for PID2024-156285NB-C41, -C42- C43, funded by MICIU/AEI/10.13039/501100011033 and by FEDER, EU, and for CNS2023-144099; Generalitat Valenciana for CIDEGENT/2020/049, CIDEGENT/2021/23, CIDEIG/2023/20, CIPROM/2023/51 and INNVA1/2024/110 (IVACE+i), Fundaci\'{o}n Bancaria La Caixa (ID 100010434), for LCF/BQ/PI25/12100025, and 2024 Leonardo Grant from BBVA Foundation, Spain;
%UAE
Khalifa University internal grants (ESIG-2023-008, RIG-2023-070 and RIG-2024-047), United Arab Emirates;
%UK
The European Union's Horizon 2020 Research and Innovation Programme (ChETEC-INFRA - Project no. 101008324).
%MC
% disclaimer

Views and opinions expressed are those of the author(s) only and do not necessarily reflect those of the European Union or the European Research Council. Neither the European Union nor the granting authority can be held responsible for them.

\bibliographystyle{apsrev4-2}
\bibliography{references}

\end{document}

%% file: authors_new.tex
\collaboration{KM3NeT Collaboration}
\email[Corresponding author: ]{km3net-pc@km3net.de}
\noaffiliation
\author{O.~Adriani\,\orcidlink{0000-0002-3592-0654}}
\affiliation{INFN, Sezione di Firenze, via Sansone 1, Sesto Fiorentino, 50019 Italy}
\affiliation{Universit{\`a} di Firenze, Dipartimento di Fisica e Astronomia, via Sansone 1, Sesto Fiorentino, 50019 Italy}
\author{J.\,A.~Aguilar\,\orcidlink{0000-0003-2252-9514}}
\affiliation{Universit{\'e}~Libre~de~Bruxelles,~Science~Faculty~CP230,~B-1050 Brussels,~Belgium}
\author{A.~Albert}
\affiliation{Universit{\'e}~de~Strasbourg,~CNRS,~IPHC~UMR~7178,~F-67000~Strasbourg,~France}
\affiliation{Universit{\'e} de Haute Alsace, rue des Fr{\`e}res Lumi{\`e}re, 68093 Mulhouse Cedex, France}
\author{A.\,R.~Alhebsi\,\orcidlink{0009-0002-7320-7638}}
\affiliation{Khalifa University of Science and Technology, Department of Physics, PO Box 127788, Abu Dhabi, United Arab Emirates}
\author{S.~Alshalloudi\,\orcidlink{0009-0000-6757-7224}}
\affiliation{Khalifa University of Science and Technology, Department of Physics, PO Box 127788, Abu Dhabi, United Arab Emirates}
\author{F.~Ameli}
\affiliation{INFN, Sezione di Roma, Piazzale Aldo Moro, 2 - c/o Dipartimento di Fisica, Edificio, G.Marconi, Roma, 00185 Italy}
\author{M.~Andre}
\affiliation{Universitat Polit{\`e}cnica de Catalunya, Laboratori d'Aplicacions Bioac{\'u}stiques, Centre Tecnol{\`o}gic de Vilanova i la Geltr{\'u}, Avda. Rambla Exposici{\'o}, s/n, Vilanova i la Geltr{\'u}, 08800 Spain}
\author{L.~Aphecetche\,\orcidlink{0000-0001-7662-3878}}
\affiliation{Subatech, IMT Atlantique, IN2P3-CNRS, Nantes Universit{\'e}, 4 rue Alfred Kastler - La Chantrerie, Nantes, BP 20722 44307 France}
\author{M. Ardid\,\orcidlink{0000-0002-3199-594X}}
\affiliation{Universitat Polit{\`e}cnica de Val{\`e}ncia, Instituto de Investigaci{\'o}n para la Gesti{\'o}n Integrada de las Zonas Costeras, C/ Paranimf, 1, Gandia, 46730 Spain}
\author{S. Ardid\,\orcidlink{0000-0003-4821-6655}}
\affiliation{Universitat Polit{\`e}cnica de Val{\`e}ncia, Instituto de Investigaci{\'o}n para la Gesti{\'o}n Integrada de las Zonas Costeras, C/ Paranimf, 1, Gandia, 46730 Spain}
% ...
